\documentclass[twocolumn,showpacs,showkeys,nofootinbib,superscriptaddress]{revtex4}
\usepackage{graphicx}
\usepackage{amsmath,amsfonts,amssymb,bm}
\begin{document}

\title{Muon transfer from deuterium to helium}

\author{M.~Augsburger} 
\affiliation{Department of Physics, University of Fribourg, CH--1700
Fribourg, Switzerland}

\author{P.~Ackerbauer}
\affiliation{Institute for Medium Energy Physics, Austrian Academy of
Sciences, A--1090 Vienna, Austria}

\author{W.H.~Breunlich}
\affiliation{Institute for Medium Energy Physics, Austrian Academy of
Sciences, A--1090 Vienna, Austria}

\author{M.~Cargnelli}
\affiliation{Institute for Medium Energy Physics, Austrian Academy of
Sciences, A--1090 Vienna, Austria}

\author{D.~Chatellard}
\affiliation{Department of Physics, University of Fribourg, CH--1700
Fribourg, Switzerland}

\author{J.-P.~Egger}
\altaffiliation{Corresponding author}
\email{Jean-Pierre.Egger@unine.ch}
\affiliation{Institut de Physique de l'Universit\'e, CH--2000
Neuch\^atel, Switzerland}

\author{B.~Gartner}
\affiliation{Institute for Medium Energy Physics, Austrian Academy of
Sciences, A--1090 Vienna, Austria}

\author{F.J.~Hartmann} 
\affiliation{Physik--Department, Technische Universit\"{a}t
M\"{u}nchen, D--85747 Garching, Germany}

\author{O.~Huot}
\affiliation{Department of Physics, University of Fribourg, CH--1700
Fribourg, Switzerland}

\author{R.~Jacot--Guillarmod}
\affiliation{Department of Physics, University of Fribourg, CH--1700
Fribourg, Switzerland}

\author{P.~Kammel}
\affiliation{Institute for Medium Energy Physics, Austrian Academy of
Sciences, A--1090 Vienna, Austria}

\author{R.~King}
\affiliation{Institute for Medium Energy Physics, Austrian Academy of
Sciences, A--1090 Vienna, Austria}

\author{P.~Knowles}
\affiliation{Department of Physics, University of Fribourg, CH--1700
Fribourg, Switzerland}

\author{A.~Kosak}
\affiliation{Physik--Department, Technische Universit\"{a}t
M\"{u}nchen, D--85747 Garching, Germany}

\author{B.~Lauss}
\affiliation{Institute for Medium Energy Physics, Austrian Academy of
Sciences, A--1090 Vienna, Austria}

\author{J.~Marton}
\affiliation{Institute for Medium Energy Physics, Austrian Academy of
Sciences, A--1090 Vienna, Austria}

\author{M.~M\"{u}hlbauer}
\affiliation{Physik--Department, Technische Universit\"{a}t
M\"{u}nchen, D--85747 Garching, Germany}

\author{F.~Mulhauser}
\altaffiliation{Corresponding author}
\email{Francoise.Mulhauser@unifr.ch}
\affiliation{Department of Physics, University of Fribourg, CH--1700
Fribourg, Switzerland}

\author{C.~Petitjean}
\affiliation{Paul Scherrer Institute, CH--5232 Villigen, Switzerland}

\author{W.~Prymas}
\affiliation{Institute for Medium Energy Physics, Austrian Academy of
Sciences, A--1090 Vienna, Austria}

\author{L.A.~Schaller} 
\affiliation{Department of Physics, University of Fribourg, CH--1700
Fribourg, Switzerland}

\author{L.~Schellenberg}
\altaffiliation{deceased}
{\affiliation{Department of Physics, University of Fribourg, CH--1700
Fribourg, Switzerland}

\author{H.~Schneuwly}
\affiliation{Department of Physics, University of Fribourg, CH--1700
Fribourg, Switzerland}

\author{S.~Tresch}
\affiliation{Department of Physics, University of Fribourg, CH--1700
Fribourg, Switzerland}

\author{T.~von Egidy}
\affiliation{Physik--Department, Technische Universit\"{a}t
M\"{u}nchen, D--85747 Garching, Germany}

\author{J.~Zmeskal}
\affiliation{Institute for Medium Energy Physics, Austrian Academy of
Sciences, A--1090 Vienna, Austria}

\date{\today}

\begin{abstract}

We report on an experiment at the Paul Scherrer Institute, Villigen,
Switzerland measuring x~rays from muon transfer from deuterium to
helium.
Both the ground state transfer via the exotic $(\mathrm{d}\mu
{}^{3,4}\mathrm{He})^{*}$ molecules and the excited state transfer
from $(\mu\mathrm{d})^*$ were measured.
The use of CCD detectors allowed x~rays from 1.5~keV to 11~keV to be
detected with sufficient energy resolution to separate the transitions
to different final states in both deuterium and helium.
The x--ray peaks of the $(\mathrm{d}\mu {}^{3}\mathrm{He})^{*}$ and
$(\mathrm{d}\mu {}^{4}\mathrm{He})^{*}$ molecules were
measured with good statistics.
For the $\mathrm{D}_2+{}^{3}\mathrm{He}$ mixture, the peak has its
maximum at $\mathrm{E}_{\mathrm{d} \mu {}^{3}\mathrm{He}} = 6768 \pm
12$~eV with FWHM $\Gamma_{\mathrm{d} \mu {}^{3} \mathrm{He}} = 863 \pm
10$~eV\@.
Furthermore the radiative branching ratio was found to be
$\kappa_{\mathrm{d}\mu{}^3\mathrm{He}} = 0.301 \pm 0.061$.
For the $\mathrm{D}_2+{}^{4}\mathrm{He}$ mixture, the maximum of the
peak lies at $\mathrm{E}_{\mathrm{d} \mu {}^{4} \mathrm{He}} = 6831
\pm 8$~eV and the FWHM is $\Gamma_{\mathrm{d} \mu {}^{4} \mathrm{He}}
= 856 \pm 10$~eV\@.
The radiative branching ratio is
$\kappa_{\mathrm{d}\mu{}^4\mathrm{He}} = 0.636 \pm 0.097$.
The excited state transfer is limited by the probability to reach the
deuterium ground state, $\mathrm{q}_{1s}$.
This coefficient was determined for both mixtures:
$\mathrm{q}^{{}^{3}\mathrm{He}}_{1s} = 68.9 \pm 2.7$\% and
$\mathrm{q}^{{}^{4}\mathrm{He}}_{1s} = 90.1 \pm 1.5$\%.
\end{abstract}


\maketitle

\section{Introduction}
\label{sec:introduction}

Muon transfer from hydrogen to helium is a loss channel in muon
catalyzed fusion ($\mu$CF), the muon induced fusion of hydrogen
isotope nuclei~\cite{breun89}.
In the $\mu$CF cycle, where in favorable cases a negative muon can
catalyze up to 200~fusions, muon transfer to helium limits the fusion
yield.
Muon transfer from hydrogen to helium can happen during the cascade in
muonic hydrogen (excited state transfer) or from the muonic hydrogen
1s ground state through the formation of an excited, metastable
hydrogen--helium molecule $(\mathrm{h}\mu\mathrm{He})^{*}$ (h = proton
p, deuteron d, or triton t, and He $={}^{3}\mathrm{He}$ or
$^{4}\mathrm{He}$), a reaction first proposed by
Aristov~\cite{arist81}.
These molecules decay from the excited state to the unbound ground
state mostly by x--ray emission ($\mathrm{E}_{x} \sim 6.8$~keV).
Auger electron emission and $\mathrm{h}\mu\mathrm{He}$ breakup
are also possible.
The scheme of the principal transfer and decay processes is presented
in Fig.~\ref{fig:diagram}.
The muon entering a deuterium--helium mixture may be captured either
by deuterium (with probability $W_{\mathrm{d}}$) or by helium (with
probability $W_{\mathrm{He}}$) via direct capture.
The two vertical arrows indicate the cascade of the muon to the 1s
ground state.
The $\mathrm{q}_{1s}^{\mathrm{He}}$ represents the probability for the
$\mu\mathrm{d}^{*}$ to reach the ground state in the presence of
helium.
Excited state transfer is shown by the upper horizontal arrow.
Ground state transfer is shown with a rate $\lambda_{\mathrm{d} \mu
\mathrm{He}}$ via the $(\mathrm{d}\mu\mathrm{He})^{*}$ molecule.
The $(\mathrm{d}\mu\mathrm{He})^{*}$ molecule decay channels are shown
with rates $\lambda_{e}$ for the Auger decay, $\lambda_{\gamma}$ for
the x--ray channel, and $\lambda_{p}$ for the break up channel,
respectively.

\begin{figure}[b]
\begin{center}
\thinlines
\centerline{\setlength{\unitlength}{1mm}}
\footnotesize
\begin{picture}(220,160)(0,0)
\put (105,155){$\mu$$^-$}
\put (20,120){\circle{25}}
\put (13,118){$\mu\mathrm{d}^{*}$}
\put (200,120){\circle{30}}
\put (191,118){$\mu\mathrm{He}^{*}$}
\put (20,30){\circle{25}}
\put (11,28){$\mu\mathrm{d}_{1s}$}
\put (200,30){\circle{30}}
\put (189,28){$\mu\mathrm{He}_{1s}$}
\put (110,30){\oval(40,20)}
\put (91.5,28){($\mathrm{d}\mu\mathrm{He})^*$}
\put (32.5,120) {\vector(1,0){151}}
\put (60,125){excited state transfer}
\put (32.5,30) {\vector(1,0){57.5}}
\put (50,35){$\lambda_{\mathrm{d} \mu \mathrm{He}}$}
\put (130,30){\vector(1,0){53}}
\put (152,33.5){$\lambda_{\gamma}$}
\put (129,35){\line(2,1){26}}
\put (155,48){\vector(2,-1){26}}
\put (152,50){$\lambda_{p}$}
\put (129,25){\line(2,-1){26}}
\put (155,12){\vector(2,1){26}}
\put (152,18.5){$\lambda_{e}$}
\put (20,107.5){\vector(0,-1){65}}
\put (200,103){\vector(0,-1){57.5}}
\put (100,155){\vector(-3,-1){73}}
\put (115,155){\vector(3,-1){73}}
\put (60,150){$W_{\mathrm{d}}$}
\put (145,150){$W_{\mathrm{He}}$}
\put (5,75){$\mathrm{q}_{1s}^{\mathrm{He}}$}
\end{picture}
\end{center}
\caption{Scheme of the main processes induced by a $\mu^{-}$ in a
  binary--gas mixture of deuterium and helium. The fusion reactions
  are not drawn. The symbols are defined in the text.}
\label{fig:diagram}
\end{figure}

The energies and widths of the molecular states have been
characterized by measuring the x--ray energy spectra.
The most precise experiment on $(\mathrm{p}\mu\mathrm{He})^{*}$,
with its intrinsically low x--ray yield,
was carried out by our collaboration~\cite{tresc98c}.
The $(\mathrm{d}\mu\mathrm{He})^{*}$ molecules were also studied by
our collaboration~\cite{gartn00} and recently an experiment was
performed on $(\mathrm{t}\mu\mathrm{He})^{*}$~\cite{matsu02}.
In those publications, earlier less precise experiments were
referenced and discussed in detail.
Our precision of $\sim0.2$\% for the energies and $\sim1.2$\% for the
widths of the $(\mathrm{d}\mu\mathrm{He})^{*}$ molecular deexcitations
make detailed comparisons with calculations possible.
Precise results on the excited state transfer probabilities were also
obtained.
The combined use of results obtained with standard detectors and
CCD techniques allowed us to determine the
radiative branching ratio of the $(\mathrm{d}\mu\mathrm{He})^{*}$
molecules, a value which has been of considerable theoretical interest
in recent years due to its direct and unique connection to the wave
function overlap in the muonic molecule~\cite{kinox93}.
The experimental challenges in obtaining the results were overcome
thanks to months of beam time, and the use of large area Charge
Coupled Device (CCD) x--ray detectors, as well as germanium
detectors~\cite{gartn00}.
A more detailed description of the present work can be found in
Augsburger's thesis~\cite{augsbphd}.

\section{Experiment}
\label{sec:experiment}

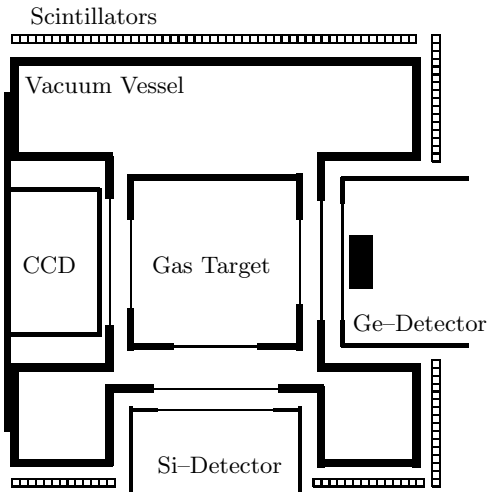
\begin{figure}[b]
\begin{center}
\thicklines
\centerline{\setlength{\unitlength}{1mm}}
\small
\begin{picture}(200,200)(0,0)
%vacuum vessel
\linethickness{1.0mm}
\put (22.5,24){\line(1,0){38.5}}
\put (60,23){\line(0,1){30.5}}
\put (59,52){\line(1,0){17.5}}
\put (124,52){\line(1,0){17}}
\put (140,53){\line(0,-1){30.5}}
\put (139,24){\line(1,0){38}}
\put (176,22.5){\line(0,1){38.5}}
\put (177.5,60){\line(-1,0){39}}
\put (140,60){\line(0,1){18}}
\put (140,123.5){\line(0,1){17.8}}
\put (140,140){\line(1,0){37.5}}
\put (176,140){\line(0,1){37.5}}
\put (176,176){\line(-1,0){153.5}}
\put (24,176){\line(0,-1){37.5}}
\put (24,140){\line(1,0){37.2}}
\put (60,140){\line(0,-1){16}}
\put (60,76){\line(0,-1){17.5}}
\put (60,60){\line(-1,0){37.2}}
\put (24,60){\line(0,-1){36}}
\thinlines
\put (76,52){\line(1,0){48}}
\put (140,76){\line(0,1){48}}
\put (60,124){\line(0,-1){48}}
% gascell
\linethickness{0.8mm}
\put (67,68){\line(1,0){17}}
\put (116,68){\line(1,0){17.}}
\put (132,66.7){\line(0,1){17.2}}
\put (132,116){\line(0,1){17.3}}
\put (133.,132){\line(-1,0){66}}
\put (68,133.2){\line(0,-1){17.2}}
\put (68,82.5){\line(0,-1){15.8}}
\thinlines
\put (84,68){\line(1,0){32}}
\put (132,84){\line(0,1){32}}
\put (68,116){\line(0,-1){34}}
% Ge 0.25 cm3
\linethickness{0.5mm}
\put (147.5,68.4){\line(1,0){48}}
\put (148,68){\line(0,1){10}}
\put (148,122){\line(0,1){10}}
\put (147.5,131.8){\line(1,0){48}}
\put (152,74){Ge--Detector}
\linethickness{3mm}
\put (155,90){\line(0,1){20}}
\thinlines
\put (148,78){\line(0,1){44}}
%Si 1cm3
\linethickness{0.5mm}
\put (68,45.3){\line(0,-1){32}}
\put (68,44){\line(1,0){10}}
\put (122,44){\line(1,0){10}}
\put (132,45.3){\line(0,-1){32}}
\put (78,20){Si--Detector}
\thinlines
\put (68,44){\line(1,0){64}}
% CCD
\linethickness{0.8mm}
\put (21.5,164){\line(0,-1){128}}
\linethickness{0.5mm}
\put (21.5,127.5){\line(1,0){34.8}}
\put (21.5,72.7){\line(1,0){34.8}}
\put (56.1,72){\line(0,1){56}}
\put (27,96){CCD}
%description
\put (28,164){Vacuum Vessel}
\put (76,96){Gas Target}
% scintillatoren
\thinlines
\put (30,190){Scintillators}
%\up
\multiput (176,186)(0,-3){2}{\line(-1,0){153}}
\multiput(176,186)(-3,0){52}{\line(0,-1){3}}
%\down
\multiput (179,15)(0,3){2}{\line(-1,0){42}}
\multiput (23,15)(0,3){2}{\line(1,0){39}}
\multiput(179,18)(-3,0){15}{\line(0,-1){3}}
\multiput(62,18)(-3,0){14}{\line(0,-1){3}}
%\side
\multiput (185,186)(-3,0){2}{\line(0,-1){48}}
\multiput (182,186)(0,-3){17}{\line(1,0){3}}
\multiput (185,15)(-3,0){2}{\line(0,1){48}}
\multiput (182,15)(0,3){17}{\line(1,0){3}}
\end{picture}
\end{center}
\caption{Schematic target setup as viewed by the entering muons.  The
  thin lines represent target and detector windows.  The drawing is
  not to scale.}
\label{fig:target}
\end{figure}

The experiment was performed at the $\mu$E4 channel at the Paul
Scherrer Institute (PSI), Villigen, Switzerland.
Setup, Ge and Si(Li) detector, gas handling and target conditions can
be taken from Tresch~\cite{tresc98c} and Gartner~\cite{gartn00}.
Figure~\ref{fig:target} shows the target setup with the detectors.
Detailed information about the large area CCD x--ray detectors is
found in Tresch~\cite{tresc98c}.

\subsection{Experimental conditions}
\label{sec:exp}

The experimental setup consisted of a gas target, Ge and Si(Li)
detectors, scintillators, and CCD detectors, as shown in
Fig.~\ref{fig:target}.
Both Tresch~\cite{tresc98c} and Gartner~\cite{gartn00} measured the
molecular formations rates $\lambda_{\mathrm{h} \mu \mathrm{He}}$ in
protium and deuterium. 
The CCDs were used by Tresch~\cite{tresc98c} for the protium
measurement.
The present work shows our results for the deuterium measurement using
CCDs.

The deuterium and related measurements, as well as the gas handling
and mixture analysis are described in great detail in
Gartner~\cite{gartn00}.
From this reference, we summarized in Table~\ref{tab:experiment} the
gas mixtures used for our analysis.
The choice of helium concentration was dictated by the goal of
Gartner~\cite{gartn00} measurement, namely the molecular formation
rates.
Due to the different theoretical values for the rates in
$\mathrm{D}_{2}+{}^{3} \mathrm{He}$ and $\mathrm{D}_{2}+{}^{4}
\mathrm{He}$, the relative concentrations of $^{3}\mathrm{He}$ and
$^{4}\mathrm{He}$ are very different.

\begin{table}[t]
\begin{ruledtabular} 
\caption{Parameters of the $\mathrm{D}_{2} + {}^{3}\mathrm{He}$ and
  $\mathrm{D}_{2} + {}^{4}\mathrm{He}$ gas mixtures.  The density
  $\phi$ is given relative to the atomic density of liquid hydrogen
  (LHD = $ 4.25 \times 10^{22}$ $\mbox{atoms}\cdot\mbox{cm}^{-3}$).}
\label{tab:experiment}
\begin{tabular}{ccccc}
Target & T & p & $\phi$ &$\mathrm{c}_{\mathrm{He}}$(atomic) \\
       & [K] & [bar] & [$10^{-3}\times$LHD] &[\%] \\ \hline
$\mathrm{D}_{2}+{}^{3}\mathrm{He}$ &$30.5\pm0.2$ & $5.58\pm0.01$ & $69.7\pm0.7$
       &$9.13\pm0.27$ \\
$\mathrm{D}_{2}+{}^{4}\mathrm{He}$ &$31.5\pm0.2$ & $5.51\pm0.01$ & $79.2\pm0.8$
       &$3.25\pm0.05$ \\
\end{tabular}
\end{ruledtabular}
\end{table}

\subsection{CCDs as low energy x--ray detectors}
\label{sec:ccd}

CCDs are excellent x--ray detectors in the energy range from
1~keV to 15~keV (details can be found in~\cite{egger99}).
In most cases, an x~ray produces charge in only a single pixel,
whereas charged particles produce cluster events or tracks with more
than one adjacent pixel hit.
The usual way to distinguish x~ray event pixels from charged particle,
neutron, and higher energy gamma--ray background is to require that
none of the eight surrounding pixels have a charge that is
considerably above the noise level.
The CCD type used for this experiment was a silicon based MOS type,
model CCD--05--20 by E2V\@.\footnote{E2V, Technologies Inc, Waterhouse
Lane, Chelmsford, Essex, CM1 2QU, England (previously EEV and
Marconi).} Each CCD chip has a size of 4.5~cm$^2$
($770\times1152$~pixels of area $22.5\times22.5 \, \mu$m$^2$) and a
depletion depth of $\sim 30 \, \mu$m.
In this experiment the energy resolution of the muonic deuterium
$\mu\mathrm{D}(2-1)$ line was 130~eV FWHM and the muonic helium
$\mu\mathrm{He}(2-1)$ transition had a resolution of 215~eV FWHM\@.

Unfortunately, our CCDs cannot be triggered so there is no timing
information.
The CCD data were read out approximately every 3~minutes by a
data-acquisition system which operated independently from the data
acquisition of the other detectors.
Therefore, we cannot normalize the collected data to the incoming
muon rate.
The results of CCD's measurements can only be analyzed using absolute
numbers.

\section{Analysis and results}
\label{sec:analysis}

\subsection{Analysis of the x--ray energy spectra}
\label{sec:ana-xray}

\begin{figure}[t]
\includegraphics[angle=90,width=0.48\textwidth]{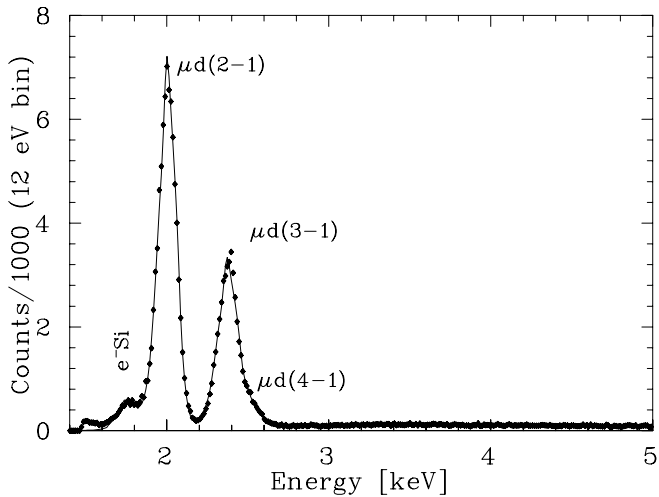}
\caption{Muonic deuterium x--ray energy spectrum. Diamonds are the
  experimental points, whereas the solid line represents the fit to
  the data.}
\label{fig:deuterium}
\end{figure}

We present in this section the different spectra obtained and explain
the fitting procedures.
Two CCDs were used.
Since each CCD half was read out separately, we have 4~sets of
measurements.
At first, the data from each half CCD were analysed separately in
order to detect any possible malfunction and to perform the energy
calibration and background reduction by single-pixel analysis.
After checking that the separate treatment of each half CCD gave
consistent results, the calibrated energy spectra were added and the
fits performed on the summed spectra.

\subsubsection{Pure element spectra}
\label{sec:d2}

The x--ray spectra from single element targets were studied in detail
to find as much information as possible about the detector response
function and target related backgrounds.  The final results required
that the entire energy range be fit at once and this was accomplished
in several steps.

\begin{figure}[t]
\includegraphics[angle=90,width=0.48\textwidth]{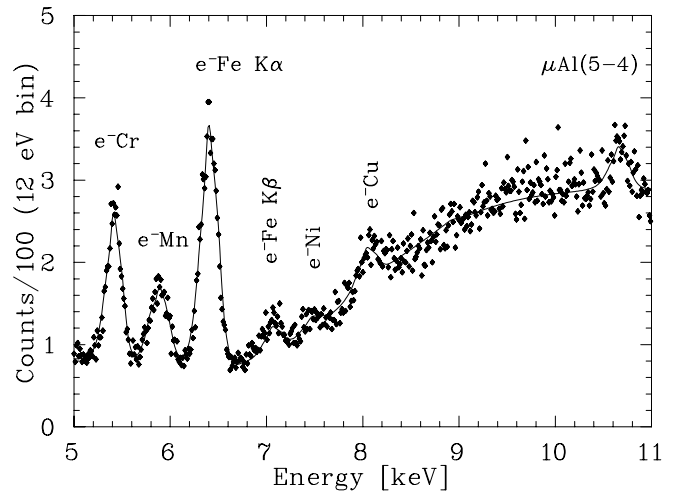}
\caption{Energy spectrum from the pure deuterium measurement (same as
  Fig.~\ref{fig:deuterium}) showing the contamination in the higher
  energy region. It is used to estimate the importance of the
  different contaminants shown (electronic Si, Cr, Mn, Fe, Ni, and Cu
  lines and muonic Al).}
\label{fig:deut-cont}
\end{figure}

\begin{figure}[b]
 \includegraphics[angle=90,width=0.48\textwidth]{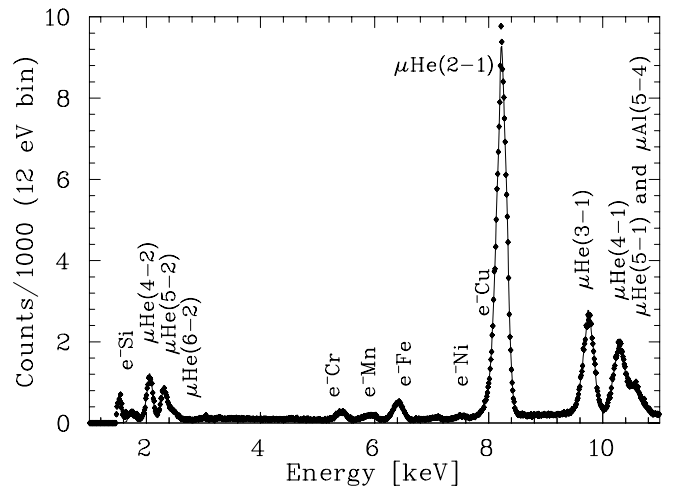}
\caption{Muonic $^{4}\mathrm{He}$ x--ray energy spectrum. The muonic
  helium Lyman series are located between 8 and 11~keV\@. The same
  contaminants as in the muonic deuterium spectrum can be seen. The
  shape of the $\mu^{4}\mathrm{He}(2-1)$ peak is used to obtain the
  standard line shape which is given in Fig.~\ref{fig:line}.}
\label{fig:helium4}
\end{figure}

The final result for the muonic deuterium x--ray spectrum is presented
in Figs.~\ref{fig:deuterium} and~\ref{fig:deut-cont}, namely the
$\mu\mathrm{d}$ x--ray transition to the 1s ground state and a series
of contaminant peaks essentially at higher energies.
Also visible are the electronic K$\alpha$ and K$\beta$ transitions of
Si (CCD), Cr, Mn, Fe, Ni, and Cu (target), although only the positions
of the K$\alpha$ peaks are indicated in Fig.~\ref{fig:deut-cont}
(except for Fe, where both lines are clearly visible).
In addition, muonic aluminium at 10.66~keV from the
$\mu\mathrm{Al}(5-4)$ transition is clearly visible.
Other muonic aluminium transitions, $\mu\mathrm{Al}(6-5)$ at 5.79~keV
and $\mu\mathrm{Al}(7-6)$ at 3.49~keV, are also present but much
weaker than the $\mu\mathrm{Al}(5-4)$ line.

The first fits were made using Gaussian peak shapes and a standard CCD
background with the goal of locating all lines and characterizing the
continuous background.
The CCD background in a high--noise environment was studied in detail
in Ref.~\cite{augsbphd}. 
The large hill starting at 7~keV seen in Fig.~\ref{fig:deut-cont} is
due essentially to energy deposited by electrons crossing the CCD\@.
The relative importance of each contaminant was estimated with the
K$\beta$/K$\alpha$ intensity ratio held fixed according to values
given in Ref.~\cite{salem74}.
Once the relative intensities of the contaminant peaks were known, the
first fits to the full spectra were carried out.
The variation of the contaminant intensities for the different spectra
were found to be small, and hence could be well parametrized.

\begin{figure}[b]
 \includegraphics[width=0.48\textwidth]{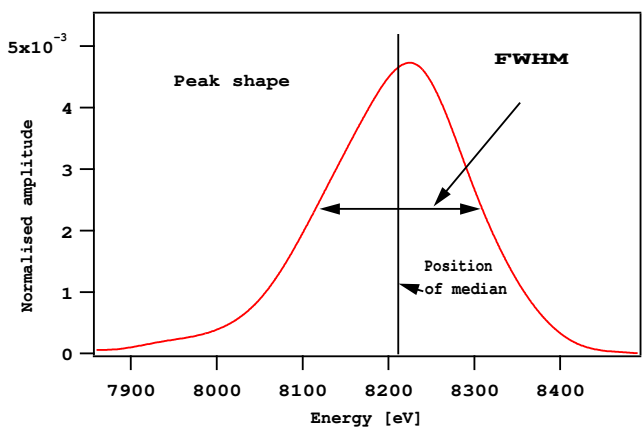}
\caption{Standard asymmetric line shape obtained from the
  $\mu{}^{4}\mathrm{He}(2-1)$ transition.  Cu contamination and
  continuous background have been subtracted.  The peak surface is
  normalized to unity. The FWHM can be adjusted.  The peak position is
  defined to be the position of the median, i.e., the center of
  gravity.}
\label{fig:line}
\end{figure}

Figure~\ref{fig:helium4} presents the spectrum for pure
$^{4}\mathrm{He}$.
The Lyman series between 8 and 11~keV and the Balmer series around
2~keV are clearly seen and the contaminant peaks are the same as in
the muonic deuterium spectrum.

\begin{figure}[t]
 \includegraphics[angle=90,width=0.46\textwidth]{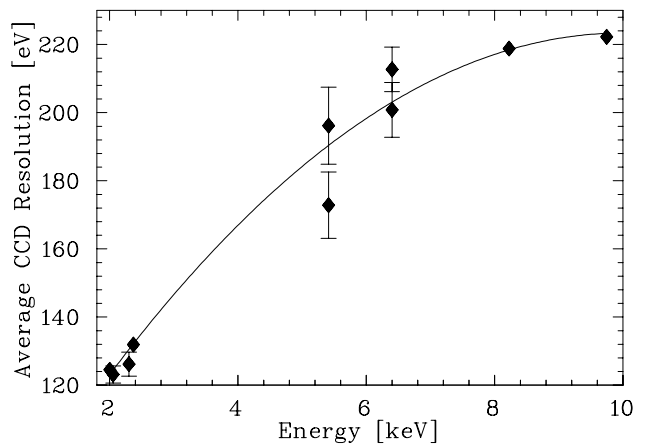}
\caption{Average FWHM energy resolution of the CCDs in eV, obtained
  after fitting the $\mu\mathrm{He}(2-1)$ transition as well as the
  electronic lines with a peak shape given in Fig.~\ref{fig:line}.
  This curve was used to constrain the FWHM of the muonic deuterium
  and helium peaks.}
\label{fig:resol}
\end{figure}

\begin{figure}[b]
 \includegraphics[angle=90,width=0.48\textwidth]{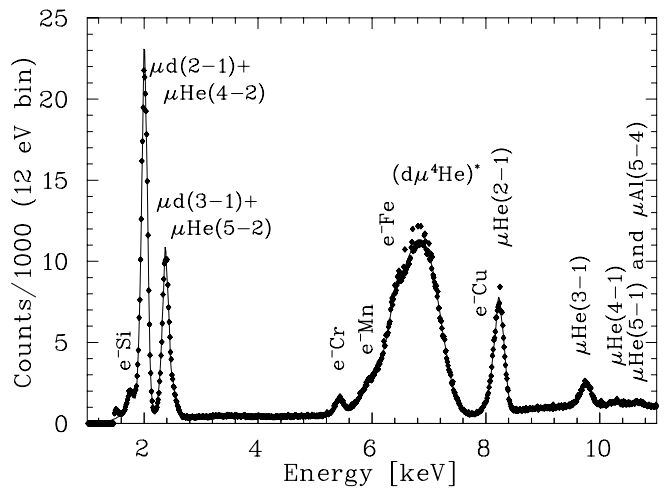}
\caption{X--ray energy spectrum of the $\mathrm{D}_{2} +
  {}^{4}\mathrm{He}$ mixture.  The large peak represents the decay of
  the $(\mathrm{d}\mu {}^{4}\mathrm{He})^{*}$ molecule via an x~ray of
  approximately 6.85~keV\@.}
\label{fig:deut-he4}
\end{figure}

To further refine the fit, the muonic helium $\mu\mathrm{He}(2-1)$
transition was examined in detail to fully understand the true CCD
response function for the line shape.
This peak was chosen, even though it contained a small copper
contamination (less than 1\%), since it has high statistics and is
well separated from the other peaks.
The small copper contribution was subtracted.
Since we could not use an analytical function for fitting the
remaining asymmetric peaks, we interpolated the asymmetric peak shape
of the $\mu\mathrm{He}(2-1)$ line, shown in Fig.~\ref{fig:line}, to
fit the data correctly.
As one can see, it looks like a Gaussian with an asymmetry on the left
side.
This same peak shape was then used to represent all other lines,
replacing the Gaussian shape, and the spectra refit.
In particular, the FWHM of each peak was obtained by using a scaling
factor from the FWHM given in Fig.~\ref{fig:line}.
In addition, all peak positions were defined by the center of gravity
(not by the maximum).
The values for the resolution of the $\mu\mathrm{He}(2-1)$ transition
as well as the electronic lines are shown in Fig.~\ref{fig:resol}.
The fitting procedure was repeated for both the muonic deuterium and
muonic helium spectra until a minimum $\chi^{2}$ was obtained.
Replacing the Gaussian line shape with the final fit function
including background parametrization and asymmetric line shapes
reduced the $\chi^{2}_{dof}$ from 5 to about 1.4 for both spectra.

\begin{figure}[t]
\includegraphics[angle=90,width=0.48\textwidth]{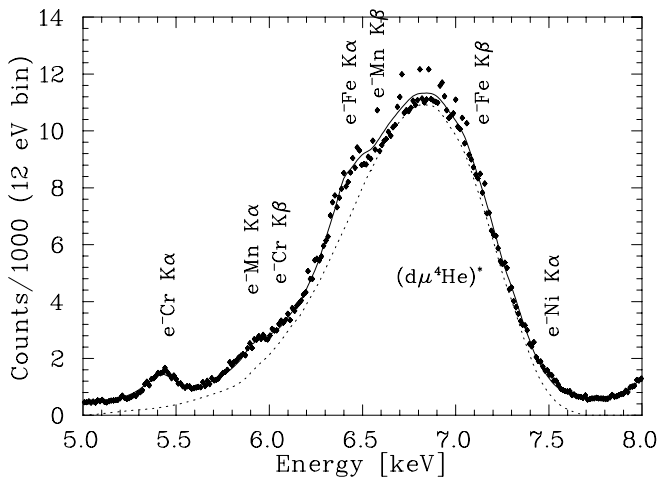}
\caption{Resulting peak shape (dotted line) of the $(\mathrm{d}\mu
  {}^{4}\mathrm{He})^{*}$ molecular x--ray after fitting contaminants
  and background (solid line).}
\label{fig:mole-he4}
\end{figure}

\subsubsection{Spectra of the $\mathrm{D}_{2} + {}^{4}\mathrm{He}$ 
and $\mathrm{D}_{2} + {}^{3}\mathrm{He}$ mixtures}
\label{sec:d2-he}

Figure~\ref{fig:deut-he4} presents the energy spectrum of the
$\mathrm{D}_{2} + {}^{4}\mathrm{He}$ mixture.
In addition to the peaks from muonic deuterium, muonic helium, and the
contaminants, a large x--ray peak from the decay of the
$(\mathrm{d}\mu {}^{4}\mathrm{He})^{*}$ molecule appears around
6.8~keV\@.
Again, the fitting procedure outlined above was used for all peaks
except the molecular peak, for which theoretical curve has been
calculated and is given in Ref.~\cite{belya97}.
The difference in shape corresponding to decays of the $J=0$ and $J=1$
state, respectively, is negligible in our case.
Hence, the calculated shape of Ref.~\cite{belya97} was taken for the
shape of this molecular peak.
The positions of the maximum is a free parameter.
We used two scaling factors to determine the amplitude and FWHM
relative to the theoretical shape.
Figure~\ref{fig:mole-he4} shows the fit of the $\mathrm{D}_2
+{}^{4}\mathrm{He}$ mixture in the region of the molecular peak (the
fit was carried out over the whole energy region, 1.6 to 11.25~keV).
The results are given in Table~\ref{tab:parameter}.

\begin{figure}[t]
\includegraphics[angle=90,width=0.48\textwidth]{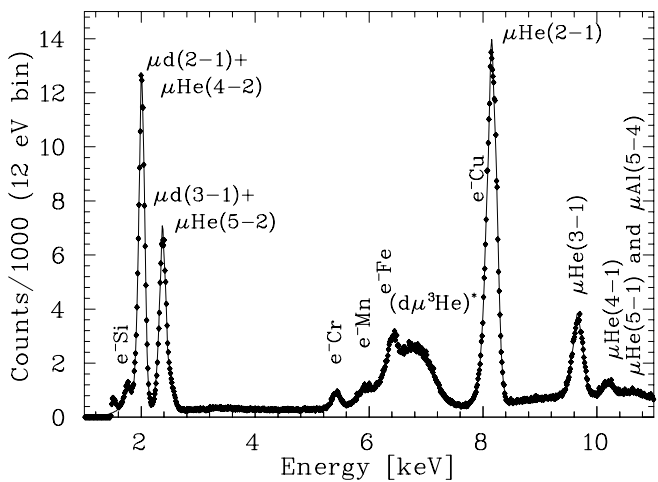}
\caption{X--ray energy spectrum of the
  $\mathrm{D}_{2}+{}^{3}\mathrm{He}$ mixture. The wide peak represents
  the decay of the $(\mathrm{d}\mu ^{3}\mathrm{He})^{*}$ molecule via
  an x~ray of about 6.8~keV\@.}
\label{fig:deut-he3}
\end{figure}

\begin{table}[b]
\begin{ruledtabular} 
\caption{Measured values of the two molecular
  $(\mathrm{d}\mu\mathrm{He})^{*}$ peaks.
  $\mathrm{E}_{\mathrm{d}\mu\mathrm{He}}$ is the energy of the peak
  maximum, $\Gamma_{\mathrm{m}}$ the measured FWHM, and
  $\Gamma_{\mathrm{d}\mu\mathrm{He}}$ the FWHM with the CCD resolution
  unfolded. $\mathrm{N}_{\mathrm{d}\mu\mathrm{He}}$ is the number of
  events in the peaks, corrected for the CCD efficiency.}
\label{tab:parameter}
\begin{tabular}{cccc} 
Value & Unit & $(\mathrm{d}\mu^{3}\mathrm{He})$ &
$(\mathrm{d}\mu^{4}\mathrm{He})$ \\ \hline
$\mathrm{E}_{\mathrm{d}\mu\mathrm{He}}$ & [eV] & $6768\pm 12$ &
$6831\pm 8$ \\
$\Gamma_{\mathrm{m}}$ & [eV] & $ 914\pm 9$ & $ 907\pm 8$ \\
$\Gamma_{\mathrm{d}\mu\mathrm{He}}$ & [eV] & $ 863\pm 10$ & $ 856\pm
10$ \\ \hline
$\mathrm{N}_{\mathrm{d} \mu \mathrm{He}}$ & &
$(411 \pm23) \times 10^3$ & $(196\pm11) \times 10^4$ \\
\end{tabular}
\end{ruledtabular}
\end{table}

Figure~\ref{fig:deut-he3} presents the spectrum of the $\mathrm{D}_2+
{}^{3}\mathrm{He}$ mixture and Fig.~\ref{fig:mole-he3} shows the same
spectrum in the region of the molecular peak.
The analysis of that spectrum was identical to the $\mathrm{D}_2 +
{}^{4}\mathrm{He}$ analysis with results also given in
Table~\ref{tab:parameter}.

\begin{figure}[t]
\includegraphics[angle=90,width=0.48\textwidth]{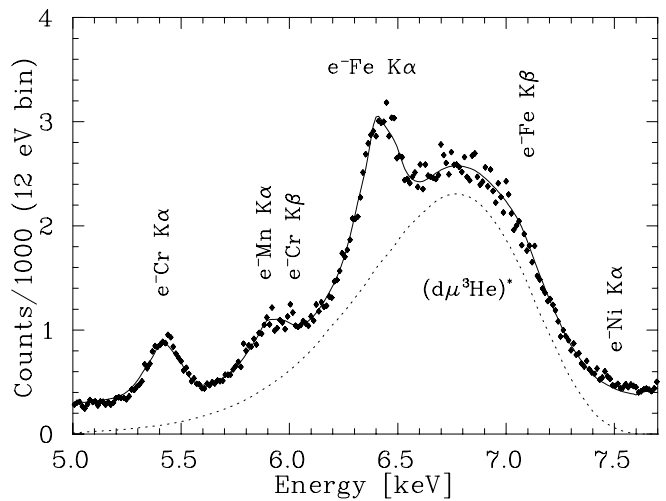}
\caption{Resulting peak shape (dotted line) of the $(\mathrm{d}\mu
  ^{3}\mathrm{He})^{*}$ molecular x--ray after fitting contaminants
  and background (solid line).}
\label{fig:mole-he3}
\end{figure}

\subsection{Relative intensities of the K series transitions in muonic
$^{4}\mathrm{He}$ }
\label{sec:intens}

The relative intensities of the K series transitions in pure muonic
$^{4}\mathrm{He}$ are given in Table~\ref{tab:relative}.
The errors include a statistical part
(fit) and a systematic part (CCD detection efficiency).
The main error comes from the CCD efficiency which is not surprising
since the fit parameters for ``CCD depletion depth'' and ``CCD window
thickness'' converge in a range of ($28\pm 2$)~$\mu$m and ($35\pm
1$)~$\mu$m respectively.
The results are compared to~\cite{tresc98c} where no isotopic effect
($^{3}\mathrm{He}$ or $^{4}\mathrm{He}$) was seen (last column of
Table~\ref{tab:relative}).
The agreement is excellent for the K$\alpha$ transition, and the
significant discrepancies of the other values are understood, since
the measurement of Tresch~\cite{tresc98c} was carried out at a lower
density, which explains the K$\beta$ decrease and the K$\gamma$
increase.
In addition, our accumulated experience with CCD background and
detection efficiency resulted in a better fit in this work but we
realize that the errors given in~\cite{tresc98c} were underestimated
with respect to the CCD efficiency correction.

\begin{table}[t]
\begin{ruledtabular} 
\caption{Relative intensities K$_{i}$/K$_{total}$ (for $i = 2,3,
  \ldots, \infty$) of the K series transitions for pure
  $^{4}\mathrm{He}$, corrected for CCD detection efficiency.  The
  errors include both statistical and CCD efficiency errors. The last
  column shows the results from Tresch~\cite{tresc98c}.}
\label{tab:relative}
\begin{tabular}{crr}
Transition & K$_{i}$/K$_{total}$ & K$_{i}$/K$_{total}$ \\
           & [\%] & [\%]~\cite{tresc98c} \\ \hline
$\mu\mathrm{He}(2-1)$ & $46.9 \pm 4.5$ &$47.0 \pm 0.2$\\
$\mu\mathrm{He}(3-1)$ & $27.9 \pm 2.8$ &$20.3 \pm 0.1$\\
$\mu\mathrm{He}(4-1)$ & $16.3 \pm 1.7$ &$19.8 \pm 0.1$\\
$\mu\mathrm{He}(5-1)$ & $ 6.2 \pm 0.7$ &$ 8.8 \pm 0.1$\\
$\mu\mathrm{He}(6-1)$ & $ 2.5 \pm 0.4$ &\\
$\mu\mathrm{He}(7-1)$ & $ 0.1 \pm 0.3$ &\\
$\mu\mathrm{He}(\infty-1)$& $ 0.1 \pm 0.1$ &$4.1\pm 1.6$ \\
\end{tabular}
\end{ruledtabular}
\end{table}

\subsection{Excited state transfer and the $\mathrm{q}_{1s}^{\mathrm{He}}$ probability}
\label{sec:excited}

The $\mathrm{q}_{1s}$ value represents the probability for a newly
formed light muonic atom to fully deexcite to the 1s state when the
muon also has the possibility of transferring directly from an excited
state to a heavier nucleus (cf. Fig.~\ref{fig:diagram}).  In binary
mixtures, the notation often includes the identity of the heavier
nucleus, $\mathrm{q}_{1s}^{\mathrm{^3He}}$ for example.

\begin{table*}[t]
\begin{ruledtabular} 
\caption{Number of events in the K~series transitions of $\mu
  \mathrm{d}$ and $\mu{}\mathrm{He}$ in a gaseous mixture of deuterium
  and $^{4}\mathrm{He}$ as well as in a gaseous mixture of deuterium
  and $^{3}\mathrm{He}$. All values are corrected for CCD detection
  efficiency.}
\label{tab:event-he4}
\begin{tabular}{c|cc|cc}
& \multicolumn{2}{c|}{$\mathrm{D}_2+{}^{4}\mathrm{He}$} &
\multicolumn{2}{c}{$\mathrm{D}_2+{}^{3}\mathrm{He}$} \\
Transition
 &\multicolumn{1}{c}{$\mu \mathrm{d}$ [$\times 10^3$] }
 &\multicolumn{1}{c|}{$\mu{}^{4}\mathrm{He}$ [$\times 10^3$]}
 &\multicolumn{1}{c}{$\mu \mathrm{d}$ [$\times 10^3$] }
 &\multicolumn{1}{c}{$\mu{}^{3}\mathrm{He}$ [$\times 10^3$]} \\
\hline
$(2-1)$ & $2915\pm173$ & $438\pm26 $
           & $1445\pm86$ & $805\pm48 $ \\
$(3-1)$ & $ 603\pm21$ & $154\pm10 $
           & $ 385\pm14$ & $287\pm19 $ \\
$(4-1)$ & $85.8\pm3.4$ & $38.6\pm2.9 $
           & $55.4\pm2.8$ & $61.9\pm4.2 $\\
$(5-1)$ & $2.3\pm 2.8$ & $ 15.9\pm3.8$
           & $1.4\pm 2.3$ & $15.4\pm1.9$\\
$(\geq6-1)$& $ 1.2\pm1.8$ & $ 10.9\pm3.9$
           & $1.0\pm1.5$ & $ 4.0\pm3.4$\\ \hline
total & $\mathrm{N}_{\mu \mathrm{d}}^{{}^4\mathrm{He}}(1s) =
3608\pm174$ & $657\pm29 $ &
$\mathrm{N}_{\mu \mathrm{d}}^{{}^3\mathrm{He}}(1s) = 1889\pm87$ &
$1174\pm52 $\\
\end{tabular}
\end{ruledtabular}
\end{table*}

We begin our analysis with the $\mathrm{D}_{2} +{}^{4}\mathrm{He}$
mixture.
The number of events in the K series transitions in $\mu \mathrm{d}$
and $\mu{}^{4}\mathrm{He}$ in the gaseous mixture of $\mathrm{D}_2 +
{}^{4}\mathrm{He}$ is given in Table~\ref{tab:event-he4}.
The sum of $\mu \mathrm{d}$ events represents the total number of $\mu
\mathrm{d}$ which reach the ground state and is called
$\mathrm{N}_{\mu \mathrm{d}}^{{}^4\mathrm{He}}(1s)$.

Part of the $^{4}\mathrm{He}$ events come from the direct capture of
the muon by helium, the other part by excited state transfer from
muonic deuterium.
It was shown in~\cite{tresc98c} that in gaseous $\mathrm{H}_2 +
{}^{4}\mathrm{He}$ mixtures, excited--state transfer proceeds only to
the levels $n=3$ and $n=2$ of $\mu^{4}\mathrm{He}$.
A detailed comparison of Figs.~\ref{fig:deut-he4} and
\ref{fig:helium4} also shows that in the $\mathrm{D}_2 +
{}^4\mathrm{He}$ mixture there is a large enhancement of the K$\alpha$
and K$\beta$ He lines over the higher transitions.
In addition, some L transitions can also be seen in
Fig.~\ref{fig:helium4}.
The fact that the L$\gamma$ transition (in Fig.~\ref{fig:deut-he4})
contains approximately $200\:000$ (efficiency corrected) events versus
only 2000 for the L$\delta$ line further confirms the above
hypothesis.
Therefore the sum of the events from the $\mu\mathrm{He}(4-1)$,
$\mu\mathrm{He}(5-1)$, and $\mu\mathrm{He}(\geq6-1)$ transitions in
$\mu{}^{4}\mathrm{He}$ is due to direct capture.
Taking this sum from Table~\ref{tab:event-he4}, one gets a measured
number of direct capture, $\mathrm{N}_{\mathrm{dc}}^{\mathrm{m}} =
(65.5\pm6.2)\times 10^3$,
where we added the errors quadratically.

In the spectrum from pure $\mu{}^{4}\mathrm{He}$ (see results in
Table~\ref{tab:relative}) the percentage sum of the
K$_{i}$/K$_{total}$ fractions for $i\geq 4$ is $25.20 \pm 1.81$\%.
The $\mathrm{N}_{\mathrm{dc}}^{\mathrm{m}}$ therefore corresponds to
25.20\% of the total number of K--series x~rays in
$\mu{}^{4}\mathrm{He}$ ($\mathrm{N}_{\mathrm{dc}}^{\mathrm{tot}}$).
Thus, we deduced the total number of direct capture events
$\mathrm{N}_{\mathrm{dc}}^{\mathrm{tot}} = (260\pm43) \times 10^3$.
This number will allow us to differentiate between direct capture and
excited state transfer events in the K$\alpha$ and K$\beta$
intensities, measured in the mixture of deuterium and
${}^{4}\mathrm{He}$.

The numbers of K$\alpha$ and K$\beta$ events occurring with the direct
capture, $\mathrm{N}_{\mathrm{dc}}^{\mathrm{K}_\alpha}$ and
$\mathrm{N}_{\mathrm{dc}}^{\mathrm{K}_\beta}$, were obtained using
the intensity ratios K$_{i}$/K$_{total}$ determined in the pure
$\mu{}^{4}\mathrm{He}$ spectrum and
$\mathrm{N}_{\mathrm{dc}}^{\mathrm{tot}}$.
The total number of $\mu{}^{4}\mathrm{He}$ K$\alpha$ and K$\beta$
events in the mixture is given in Table~\ref{tab:event-he4}.
The differences are due to excited state transfer from $\mu
\mathrm{d}^{*}$.
The number of K$\alpha$ and K$\beta$ events coming from excited state
transfer, $\mathrm{N}_{\mathrm{exc}}^{\mathrm{K}_\alpha}$ and
$\mathrm{N}_{\mathrm{exc}}^{\mathrm{K}_\beta}$, are the difference
between the first two lines of Table~\ref{tab:event-he4} and the
previously determined $\mathrm{N}_{\mathrm{dc}}^{\mathrm{K}_\alpha}$
and $\mathrm{N}_{\mathrm{dc}}^{\mathrm{K}_\beta}$.
Therefore the sum of events from excited state transfer is
$\mathrm{N}_{\mathrm{exc}}^{\mathrm{tot}} = (398\pm47)\times 10^3 $.
Now $\mathrm{q}_{1s}^{^{4}\mathrm{He}}$ can be determined by
\begin{equation}
\mathrm{q}_{1s}^{^{4}\mathrm{He}} =
\frac{\mathrm{N}_{\mu \mathrm{d}}^{{}^4\mathrm{He}}(1s)}
     {\mathrm{N}_{\mu \mathrm{d}}^{{}^4\mathrm{He}}(1s)
     + \mathrm{N}_{\mathrm{exc}}^{\mathrm{tot}}} = 90.1 \pm 1.5 \, \%
  \label{eq:3}
\end{equation}
where $\mathrm{N}_{\mu \mathrm{d}}^{{}^4\mathrm{He}}(1s)$ is the total
number of $\mu
\mathrm{d}$ Lyman x~rays in the $\mathrm{D}_{2} + {}^{4}\mathrm{He}$
mixture (see Table~\ref{tab:event-he4}).

The analysis carried out in the case of the $\mathrm{D}_{2} +
{}^{3}\mathrm{He}$ mixture was the same as in the $\mathrm{D}_{2} +
{}^{4}\mathrm{He}$ mixture with the additional hypothesis that the
muonic cascade was the same in both $\mu{}^{3}\mathrm{He}$ and
$\mu{}^{4}\mathrm{He}$.
Pure $\mu{}^{3}\mathrm{He}$ was not measured (only
$\mu{}^{4}\mathrm{He}$) for this work.
However, Tresch~\cite{tresc98c} has shown no isotopic effects between
the two gases. 

Therefore, the $\mathrm{q}_{1s}^{^{3}\mathrm{He}}$ is determined as
\begin{equation}
 \mathrm{q}_{1s}^{^{3}\mathrm{He}} =
  \frac{\mathrm{N}_{\mu \mathrm{d}}^{{}^3\mathrm{He}}(1s)}
       {\mathrm{N}_{\mu \mathrm{d}}^{{}^3\mathrm{He}}(1s)
     + \mathrm{N}_{\mathrm{exc}}^{\mathrm{tot}}} = 68.9 \pm 2.7 \, \%
\label{eq:4}
\end{equation}
where $\mathrm{N}_{\mu \mathrm{d}}^{{}^3\mathrm{He}}(1s)$ is the total
number of $\mu \mathrm{d}$ Lyman x~rays in the $\mathrm{D}_{2} +
{}^{3}\mathrm{He}$ mixture (Table~\ref{tab:event-he4}).

\subsection{Radiative branching ratio
  $\kappa_{\mathrm{d}\mu\mathrm{He}}$ of the
  $(\mathrm{d}\mu\mathrm{He})^{*}$ molecule}
\label{sec:branching}

The radiative branching ratio $\kappa_{\mathrm{d}\mu\mathrm{He}}$ for
the $(\mathrm{d}\mu\mathrm{He})^{*}$ molecular decay can be determined
the same way as in~\cite{tresc98c} for the
$(\mathrm{p}\mu\mathrm{He})^{*}$ molecule.
$\kappa_{\mathrm{d}\mu\mathrm{He}}$ is given by
\begin{equation}
\kappa_{\mathrm{d}\mu\mathrm{He}} \cdot W =
    \frac{\mathrm{N}_{\mathrm{d}\mu\mathrm{He}}}
	 {\mathrm{N}_{\mu \mathrm{d}}^{\mathrm{He}}(1s) }\, ,
\label{eq:1}
\end{equation}
where $\mathrm{N}_{\mathrm{d} \mu \mathrm{He}}$ is the number of
events in the molecular peak (see Table~\ref{tab:parameter}) and
$\mathrm{N}_{\mu \mathrm{d}}^{\mathrm{He}}(1s)$ the total number of
the $\mu \mathrm{d}$ Lyman series x~rays in the mixture.
$W$ is the probability of a $\mu\mathrm{d}_{\mathrm{1s}}$ forming a
$(\mathrm{d}\mu\mathrm{He})^{*}$ molecule and is given by the
equation:
\begin{equation}
W = \frac{\phi \mathrm{c}_{\mathrm{He}} \lambda_{\mathrm{d} \mu
\mathrm{He}}}
{\Lambda_{\mu \mathrm{d}_{1s}}}\, ,
\label{eq:2}
\end{equation}
where $\phi$ is the atomic density of the mixture, normalized to LHD,
$\mathrm{c}_\mathrm{He}$ is the helium atom proportion,
$\lambda_{\mathrm{d} \mu \mathrm{He}}$ the ground state transfer rate
from $\mu \mathrm{d}$ to He, and $\Lambda_{\mu \mathrm{d}_{1s}}$ the
disappearance rate of muons from the $(\mu \mathrm{d})_{1s}$ level.

\begin{table}[ht]
\begin{ruledtabular} 
\caption{Radiative branching ratio $\kappa_{\mathrm{d}\mu\mathrm{He}}$
  of the $(\mathrm{d}\mu{}^{3}\mathrm{He})^{*}$ and
  $(\mathrm{d}\mu{}^{4}\mathrm{He})^{*}$ molecules. The errors are
  commented in the text.}
\label{tab:branching}
\begin{tabular}{ccr}
             &$\mathrm{d}\mu{}^{3}\mathrm{He}$
&\multicolumn{1}{c}{$\mathrm{d}\mu{}^{4}\mathrm{He}$}
\\ \hline
$\lambda_{\mathrm{d} \mu \mathrm{He}}\:(10^{8}\:\mbox{s}^{-1})$
             & $1.856 \pm 0.077$ & 
$\phantom{1} 10.50 \pm 0.21$ \\
$\Lambda_{\mu \mathrm{d}_{1s}}\:(10^{6}\:\mbox{s}^{-1})$
             & $1.637 \pm 0.032 $ & $3.159 \pm 0.018$ \\
$W$ & $0.721 \pm 0.073 $ & $0.856 \pm 0.044$ \\
$\kappa_{\mathrm{d}\mu\mathrm{He}}$ & $0.301 \pm 0.061 $ & $0.636 \pm
0.097$ \\
\end{tabular}
\end{ruledtabular}
\end{table}

The so determined values for $\kappa_{\mathrm{d}\mu\mathrm{He}}$ and
$W$ are given in Table~\ref{tab:branching} for both helium isotopes.
The values for $\lambda_{\mathrm{d} \mu \mathrm{He}}$ and
$\Lambda_{\mu \mathrm{d}_{1s}}$ (in Table~\ref{tab:branching}) were
taken from Gartner~\cite{gartn00}.
While the errors for $\mathrm{N}_{\mathrm{d} \mu \mathrm{He}}$ and
$\mathrm{N}_{\mu \mathrm{d}}^{\mathrm{He}}(1s)$ include both the
statistical and systematic uncertainties, the errors in $\phi$ and
$\mathrm{c}_{\mathrm{He}}$ given in Table~\ref{tab:experiment} are
purely systematic.
The errors on $W$ and $\kappa_{\mathrm{d}\mu\mathrm{He}}$ were
calculated by normal error propagation without specifying the type of
error.

\section{Discussions}
\label{sec:discussion}

\subsection{General features of the x--ray energy spectra}
\label{sec:feature}

The spectra presented in Figs.~\ref{fig:deuterium},
\ref{fig:deut-cont}, \ref{fig:helium4}, \ref{fig:deut-he4}, and
\ref{fig:deut-he3} deserve three general comments.
First, the relative intensities of the muonic deuterium K$\alpha$ and
K$\beta$ transitions are density dependent~\cite{lauss98}.
Since the density changed between mixtures (see
Table~\ref{tab:experiment}), the muonic deuterium K$\alpha$ peak is
slightly enhanced over K$\beta$ in the $\mathrm{D}_2 +
{}^{4}\mathrm{He}$ mixture.

Second, the x--ray count rate for the $(\mathrm{d}\mu
{}^{3}\mathrm{He})^{*}$ molecule is smaller than for the
$(\mathrm{d}\mu {}^{4}\mathrm{He})^{*}$ molecule since in the
$(\mathrm{d}\mu {}^{3}\mathrm{He})^{*}$ case the two--particle breakup
channel is more prominent.
Third, the relative helium/deuterium line intensities depend on the
helium concentration (see Table~\ref{tab:experiment}).

\subsection{$(\mathrm{d}\mu {}^{3}\mathrm{He})^{*}$ and 
$(\mathrm{d}\mu {}^{4}\mathrm{He})^{*}$ molecules}
\label{sec:molecule}

\begin{table}[b]
\begin{ruledtabular} 
\caption{Theoretical and experimental energies of the maximum of the
  molecular peaks, in eV\@.  For the experimental values, we also list
  the detector type.  Our values are taken from
  Table~\ref{tab:parameter}.}
\label{tab:theory}
\begin{tabular}{ccccc} 
Theory &\multicolumn{2}{c}{$(\mathrm{d}\mu{}^{3}\mathrm{He})$}
       & \multicolumn{2}{c}{$(\mathrm{d}\mu{}^{4}\mathrm{He})$} \\
            & $J=0$ & $J=1$ & $J=0$ & $J=1$ \\ \hline
Belyaev~\cite{belya97}
            & 6766 & 6808 & 6836 & 6878 \\
Czaplinski~\cite{czapl97b}
            & 6760 & 6782 & 6836 & 6857 \\ \hline
Experiment &\multicolumn{2}{c}{$(\mathrm{d}\mu{}^{3}\mathrm{He})$}
           & \multicolumn{2}{c}{$(\mathrm{d}\mu{}^{4}\mathrm{He})$} \\
\hline
Gartner~\cite{gartn00} \\
Ge + Si(Li)&\multicolumn{2}{c}{$(6.80\pm0.03)\times10^{3}$}
           &\multicolumn{2}{c}{$(6.88\pm0.03)\times10^{3}$} \\
this work \\
CCD &\multicolumn{2}{c}{$6768 \pm 12$}&\multicolumn{2}{c}{$6831\pm 8$}
\\
\end{tabular}
\end{ruledtabular}
\end{table}

In Table~\ref{tab:theory} our CCD results for the position of the
maximum of the two molecular peaks are compared to theoretical
predictions~\cite{belya97,czapl97b} and to results~\cite{gartn00}
obtained with Ge and Si(Li) detectors.
The Ge and Si(Li) detector results seem to favor the J=1 state but the
CCD results imply a preference for transitions from $J=0$.
Even if the CCD results are more precise and the CCD statistics are
significantly higher, it is difficult to decide for $J=0$ or $J=1$
since the results of both detector types are effectively compatible
considering the errors.
What can be said unambiguously is that the CCD results are in
excellent agreement with both theoretical predictions for decay from
the $J=0$ state.
It should be stressed that both the Ge and Si(Li) and the CCD data
were taken simultaneously during the experiment.

\begin{table}[t]
\begin{ruledtabular} 
\caption{Theoretical and experimental widths (FWHM) of the molecular
  peaks in eV\@.  For the experimental values, we also list the
  detector type.  Our values are taken from
  Table~\ref{tab:parameter}. }
\label{tab:width}
\begin{tabular}{ccccc} 
Theory &\multicolumn{2}{c}{$(\mathrm{d}\mu{}^{3}\mathrm{He})$}
         &\multicolumn{2}{c}{$(\mathrm{d}\mu{}^{4}\mathrm{He})$} \\
         & $J=0$ & $J=1$ & $J=0$ & $J=1$ \\ \hline
Belyaev~\cite{belya97}
         & $861\pm3$ & $858\pm3$ & $843\pm3$ & $848\pm3$ \\
Czaplinski~\cite{czapl97b}
         & $866\pm3$ & $867\pm3$ & $854\pm3$ & $855\pm3$ \\ \hline
Experiment &\multicolumn{2}{c}{$(\mathrm{d}\mu{}^{3}\mathrm{He})$}
           &\multicolumn{2}{c}{$(\mathrm{d}\mu{}^{4}\mathrm{He})$}
\\\hline
Gartner~\cite{gartn00} \\
Ge +
Si(Li)&\multicolumn{2}{c}{$910\pm30$}&\multicolumn{2}{c}{$910\pm20$}
\\
this work \\
CCD &\multicolumn{2}{c}{$863\pm10$}&\multicolumn{2}{c}{$856\pm10$} \\
\end{tabular}
\end{ruledtabular}
\end{table}

In Table~\ref{tab:width} the CCD experimental FWHM widths for the
molecular peaks are again compared with the theoretical
predictions~\cite{belya97,czapl97b} and with the Ge and Si(Li)
data~\cite{gartn00}.
The CCD results are in very good agreement with theoretical
predictions for both $J=0$ and $J=1$ states, but distinguishing
between the two states is not possible due to the almost identical
theoretical values.
On the other hand, the results with the ``classic'' (Ge or Si(Li))
x--ray detectors are between 1.5 to 2.5~$\sigma$ away from theory.
The somewhat smaller width of the ($\mathrm{d}\mu{}^{4}\mathrm{He}$)
molecule predicted by theory is also hinted at by our CCD data.

\subsection{Radiative branching ratio $\kappa$}
\label{sec:radiative}

\begin{table}[t]
\begin{ruledtabular} 
\caption{Comparison of the theoretical and experimental radiative
  branching ratios $\kappa$ of the $(\mathrm{d}\mu\mathrm{He})^{*}$
  molecules.  Only Kravtsov~\cite{kravt93} includes all three
  disintegration channels (the others neglect the Auger channel, given
  in Fig.~\ref{fig:diagram}).}
\label{tab:comparison}
\begin{tabular}{c|c|c|c|c} 
 & &$\kappa_{\mathrm{d}\mu{}^{3}\mathrm{He}}$
&$\kappa_{\mathrm{d}\mu{}^{4}\mathrm{He}}$&$\frac{\kappa_{\mathrm{d}\mu{}^{3}\mathrm{He}}}{\kappa_{\mathrm{d}\mu{}^{4}\mathrm{He}}}$\\
\hline
Kino~\cite{kinox93} &$J=1$& 0.234 & 0.503 & 0.465 \\
Gershtein~\cite{gersh93}&$J=1$& 0.18 & 0.41 & 0.44 \\
Kravtsov~\cite{kravt93} &$J=0$& 0.31 & 0.45 & 0.69 \\
                        &$J=1$& 0.33 & 0.49 & 0.67 \\
Belyaev~\cite{belya95c}&$J=1$& 0.325 & 0.585 & 0.56 \\
Belyaev~\cite{belya96} &$J=0$& 0.364 & 0.707 & 0.51 \\
                        &$J=1$& 0.309 & 0.568 & 0.54 \\
this work& &$0.301\pm0.061$ & $0.636\pm0.097$&$0.47\pm0.17$ \\
\end{tabular}
\end{ruledtabular}
\end{table}

Table~\ref{tab:comparison} presents the different theoretical values
for the radiative branching ratio.
The calculations are those of Kino~\cite{kinox93},
Gershtein~\cite{gersh93}, Kravtsov~\cite{kravt93}, and
Belyaev~\cite{belya95c,belya96}.
Except for Kravtsov~\cite{kravt93} who includes the Auger decay
channel $\lambda_{e}$ (see Fig.~\ref{fig:diagram}), only $\lambda_{p}$
(break up) and $\lambda_{\gamma}$ (x~ray) are calculated.
The x~ray channel relates to $\kappa_{\mathrm{d}\mu\mathrm{He}}$
via the ratio
\begin{equation}
\kappa_{\mathrm{d}\mu\mathrm{He}} =
\frac{\lambda_{\gamma}}{\lambda_{p} + \lambda_{\gamma} +
  \lambda_{e}}\, .
\label{eq:5}  
\end{equation}
The different theoretical $\kappa_{\mathrm{d}\mu\mathrm{He}}$ values
are compared with our experiment in Table~\ref{tab:comparison}.
The large isotopic effect predicted by theory is seen by our
experiment which is in contradiction to the $(\mathrm{p}\mu
\mathrm{He})$ case~\cite{tresc98c}.
In general the agreement between theory and experiment is good,
however, the experimental errors are sizable.

\subsection{Ground state formation probabilities $\mathrm{q}_{1s}^{\mathrm{He}}$  }
\label{sec:excited-state}

\begin{table}[t]
\begin{ruledtabular} 
\caption{Comparison of the $\mathrm{q}_{1s}^{\mathrm{He}}$ probability
  in hydrogen--helium (Tresch~\cite{tresc98c}) and in
  deuterium--helium (this experiment) mixtures.}
\label{tab:q1s}
\begin{tabular}{c|cccc} 
Mixture&$\mathrm{H}_2 +
{}^{3}\mathrm{He}$&$\mathrm{H}_2+{}^{4}\mathrm{He}$
&$\mathrm{D}_2 + {}^{3}\mathrm{He}$
&$\mathrm{D}_{2} + {}^{4}\mathrm{He}$ \\ \hline
$\mathrm{q}_{1s}^{\mathrm{He}}$
[\%]&$50\pm10$&$65\pm10$&$68.9\pm2.7$&$90.1\pm1.5$\\
\end{tabular}
\end{ruledtabular}
\end{table}

The meaning of $\mathrm{q}_{1s}^{\mathrm{He}}$ has been described in
Section~\ref{sec:excited}.
Our results for the deuterium--helium mixtures, Eqs.~(\ref{eq:3}) and
(\ref{eq:4}), are listed together with those for the hydrogen--helium
mixtures~\cite{tresc98c} in Table~\ref{tab:q1s}.
It is interesting to note that in both cases (hydrogen or deuterium)
$\mathrm{q}_{1s}^{\mathrm{He}}$ is smaller for ${}^{3}\mathrm{He}$,
and therefore, the excited state transfer is more probable for the
lighter of the two helium isotopes.
In the case of hydrogen--deuterium mixtures~\cite{lauss96},
$\mathrm{q}_{1s}^{\mathrm{He}}$ has been shown to depend on the
concentration of the components of the mixture.
The large difference seen in our case between $\mathrm{D}_{2} +
{}^{3}\mathrm{He}$ and $\mathrm{D}_{2} + {}^{4}\mathrm{He}$ is
therefore partially due to the differing helium concentrations (see
Table~\ref{tab:experiment}).
The second observation is that the $\mathrm{q}_{1s}^{\mathrm{He}}$ is
significantly larger for deuterium, a result of consequence in the
case of muon catalyzed fusion in deuterium--helium
mixtures~\cite{knowl01}.

\section{Conclusions}
\label{sec:conclusion}

The use of CCDs for low energy x--ray detection allowed for a complete
energy measurement of muonic deuterium, helium, and molecular
$(\mathrm{d}\mu \mathrm{He})$ x~rays with excellent energy resolution
and low background.
The large CCD surface resulted in an increased solid angle and
therefore in better statistics when compared to traditional Ge or
Si(Li) detectors.
Of course, results like transfer rates still need the usual x--ray
detectors since the CCDs give no timing information.
The simultaneous use of CCDs and other x--ray detectors allows for
systematic error checks of the experiment since the CCD electronics is
completely independent.
In conclusion, the addition of CCDs permitted a characterization of
all transfer parameters and some high precision results.
Only the use of CCD detectors allowed the determination of the
radiative branching ratio, a result which was long awaited by
theorists.

\begin{acknowledgments}
Financial support by the Austrian Academy of Sciences, the Austrian
Science Foundation, the Swiss Academy of Sciences, the Swiss National
Science Foundation and the Beschleunigerlaboratorium of the University
and the Technical University Munich is gratefully acknowledged.
\end{acknowledgments}

%\bibliography{mucf}

\end{document}